\documentclass[preprint,12pt]{elsarticle}

\usepackage{epsr}

\usepackage{amssymb}
\usepackage{amsmath}
\usepackage{xcolor}

\journal{Nuclear Physics B}

\begin{document}

\begin{frontmatter}



\title{An Analytical Review of Model Order Reduction Methodologies of Converter-Dominated Power Systems for Stability Studies} 


\author[label1]{Goran Grdenić}
\author[label1]{Marco Fraccaro}
\author[label1]{Josipa-Pina Milišić} 

\affiliation[label1]{organization={University of Zagreb Faculty of Electrical Engineering and Computing},
            addressline={Unska 3}, 
            postcode={10000}, 
            city={Zagreb},
            country={Croatia}}

\begin{abstract}
The rapid integration of renewable energy sources is transforming modern electric power systems into converter-dominated networks characterized by complex multi-timescale dynamics and new stability interactions over a wide frequency range. Accurate analysis of these phenomena requires high-fidelity models of transmission networks and power electronic converters, which often result in large-scale differential–algebraic systems with significant computational burden. This paper provides an analytical review of model order reduction techniques for converter-rich power systems, with particular emphasis on the key equations employed in the reduction process. Advantages and limitations of commonly used strategies, including timescale-based singular perturbation, projection-based, operator-based, data-driven, and machine-learning based methods, are discussed. The synthesis highlights key challenges related to harmonic interactions and multi-frequency dynamics, and outlines directions for the development of computationally efficient yet dynamically accurate reduced-order models for future power systems.
\end{abstract}




\end{frontmatter}



\label{sec1}

We are currently witnessing a profound transformation of the electric power sector, driven primarily by the large-scale integration of renewable energy sources as a key instrument in global efforts to mitigate climate change. The electricity sector plays a central role in this transition due to the ongoing electrification of other domains \cite{EU_RED_2023}. 

\begin{figure}[hbt!]
	\centering
	\includegraphics[scale=0.5]{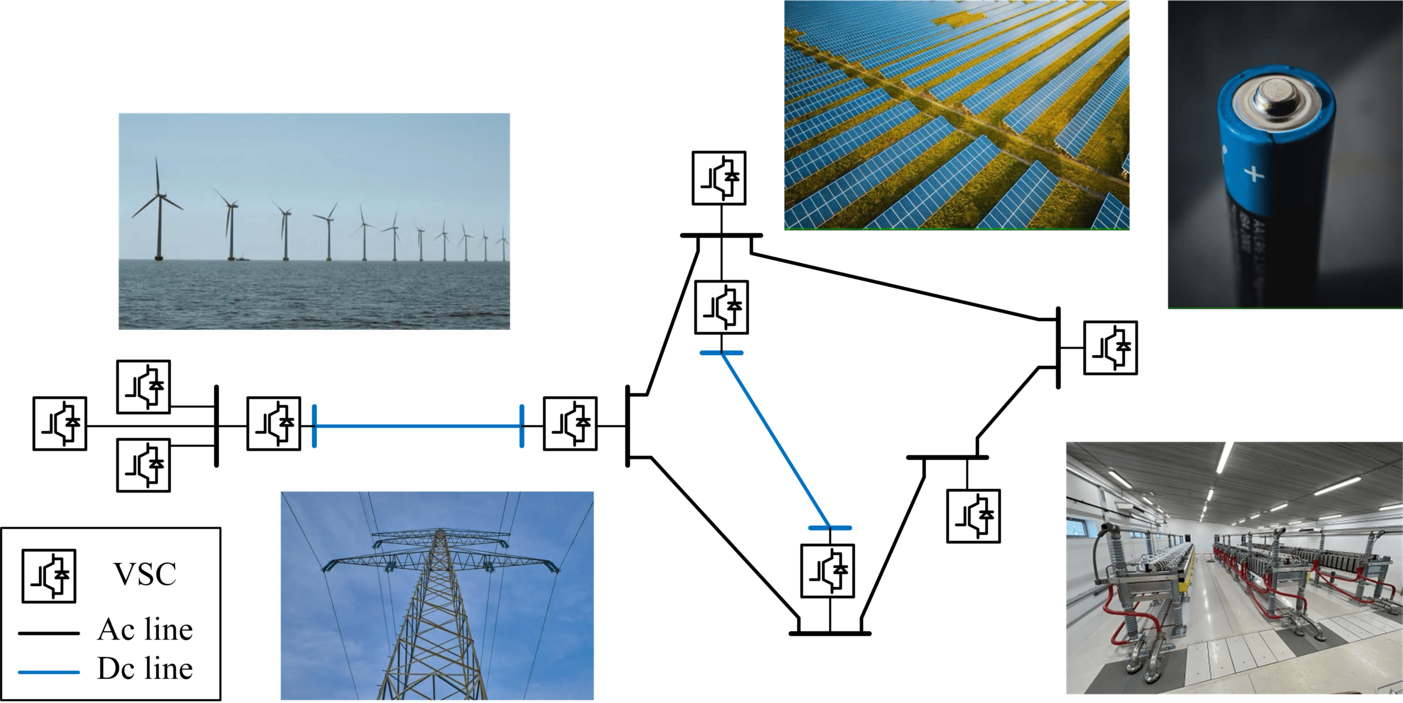}
	\caption{Renewable-based power systems}
	\label{fig:figure1}
\end{figure}

The large-scale integration of renewable energy sources is closely associated with the extensive deployment of power electronic devices (PEDs). In particular, transistor-based voltage-source converters (VSCs) have become key enabling components of modern power systems. They are widely employed in wind power plants, PVs, HVDC, FACTS, and battery energy storage systems, as illustrated in Fig. \ref{fig:figure1}.

\begin{figure}[hbt!]
	\centering
	\includegraphics[scale=0.5]{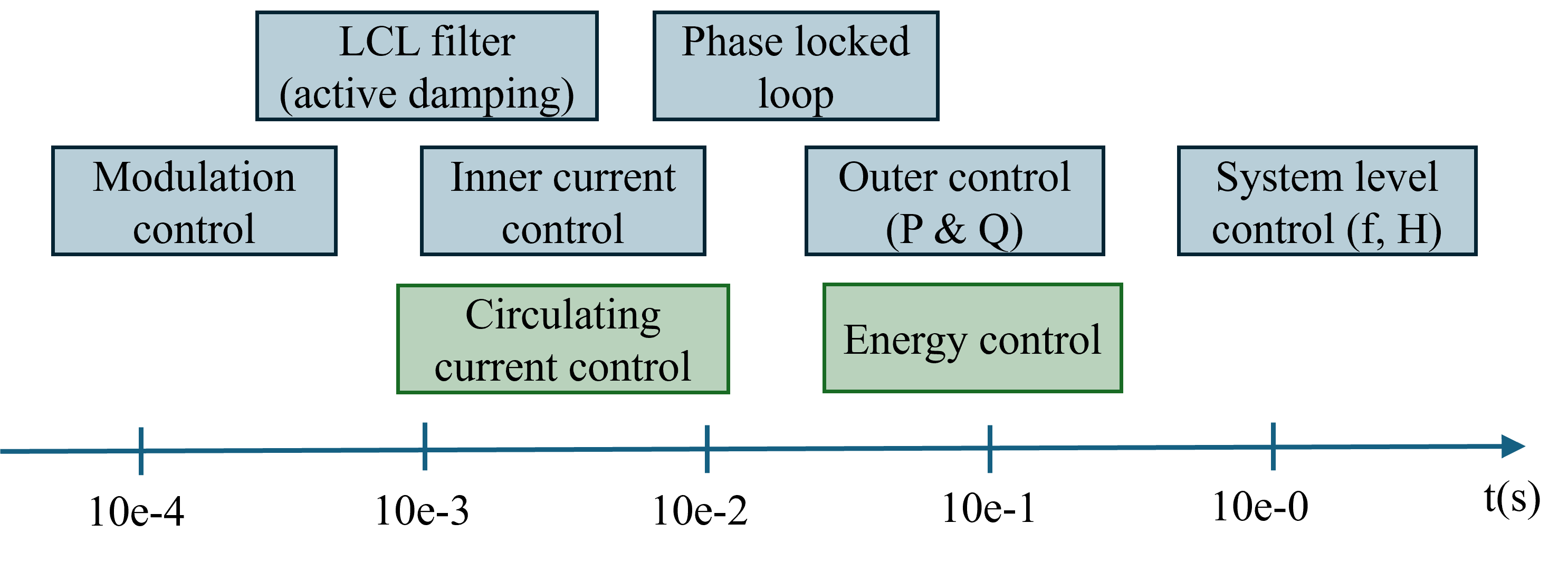}
	\caption{VSC controls and the associated time constants}
	\label{fig:figure2}
\end{figure}

Two main advantages characterize VSC devices: fast dynamic response and capability to control active and reactive power independently. The most common VSC control loops and the associated time constants are depicted in Fig. \ref{fig:figure2}. Converter control generally consists of nested control loops, including switching level modulation, inner current control (ICC), outer control (e.g., active and reactive power, AC and DC-side voltage), system level control (e.g., frequency or virtual inertia in grid-forming converters), phase locked loop (PLL) in grid-following converters, active damping for LCL filter oscillations. Additionally, modular multilevel converters (MMCs) also incorporate circulating current and energy control. The corresponding time constants span several orders of magnitude, ranging from hundreds of microseconds at the switching level to seconds at the system level control.

The increasing penetration of PEDs has introduced new forms of stability interactions in modern power systems \cite{Hatziargyriou2021, https://doi.org/10.1049/gtd2.12430}. Converter control loops can interact both with other converters and with various grid components across a wide frequency spectrum. When converters interact with power-system elements characterized by slow dynamic responses -- such as electromechanical dynamics or synchronous machines control systems -- the resulting phenomena are referred to as slow-interaction converter-driven stability.

In contrast, fast-interaction converter-driven stability arises when converters interact with components exhibiting fast dynamics, including the control systems of other PEDs and transmission line electromagnetic dynamics. In the literature, these effects are also known as electromagnetic or harmonic stability \cite{BayoSalas2018}. Field observations confirm that resonant interactions may occur in the kilohertz range \cite{Zou2018, Dennetiere2019}. An overview of real-life events, arising over subsynchronous and supersynchronous frequency ranges, can be found in \cite{DERUA2023113431} and \cite{9740416}.

Accurate identification and mitigation of such interactions require modeling frameworks with sufficient frequency-domain resolution \cite{10922127}. Conventional phasor-based network representations, although widely used for electromechanical studies, may lead to misleading stability conclusions in converter-dominated systems \cite{Amin2015,Grdenic2020}. Similarly, the commonly adopted $\pi$-equivalent transmission-line model has limited frequency validity. Therefore, the analysis of higher-frequency phenomena necessitates more advanced line representations, such as distributed-parameter and frequency-dependent models.

Converter models exhibit similar limitations. Two-level VSC averaged models formulated in the synchronous $dq$ reference frame are typically valid only up to approximately one half of the switching frequency \cite{Wang2019}. To capture switching-frequency sideband oscillations and related electromagnetic interactions, frequency-lifted modeling approaches such as dynamic-phasor or harmonic state-space formulations are required. However, such detailed models significantly increase system dimensionality. Transmission-line and converter representations that retain higher-frequency dynamics often involve tens of state variables, and the overall model order grows with the upper frequency bound considered. In large-scale systems containing numerous transmission elements and PE interfaces, the resulting dynamical models can therefore become extremely high-dimensional. At the same time, modeling all system components in full frequency detail is not always necessary for reliable stability assessment.

Converter-dominated power systems are typically operated closer to their stability limits than traditional synchronous-generator (SG)-based systems. In addition to the variability of renewable generation, limited possibilities for expanding transmission capacity in densely populated regions further constrain system operation. Since electric power systems represent critical infrastructure, ensuring dynamic security under all operating conditions is of paramount importance. This requirement necessitates frequent and computationally efficient stability assessments.

When detailed multi-timescale models are employed, stability analysis becomes computationally demanding. Time-domain simulations require very small integration time steps to resolve fast converter and electromagnetic dynamics while simultaneously covering long time horizons associated with slower processes. Frequency-domain techniques such as modal analysis lead to high-dimensional state-space representations, making eigenvalue computation challenging for large systems.

For these reasons, the development of reduced-order modeling approaches for converter-dominated power systems is essential for enabling safe and reliable operation. Such models aim to retain the dominant dynamic characteristics relevant for stability assessment while substantially reducing computational complexity.

This study area has attracted significant interest from researchers developing efficient model order reduction (MOR) methodologies for converter-dominated power systems. An extensive review of various techniques was done in \cite{Aththanayake2024PowerSystemReduction} and \cite{review}. Different from the previous work, the main objective of this paper is twofold. First, the focus of the review is narrowed to reduced-order modeling for stability studies applicable to the overall power system, and not adapted to only one power system component. Secondly, the goal is to provide an analytical overview of different approaches in which various references are not only described but also critically revisited through key expressions.

The rest of the article is organized as follows. The mathematical formulation of developing a reduced-order model, together with challenges associated with the multi-timescale nature of PED-based power systems, is presented in Section \ref{sec.MOR}. Sections \ref{sec.TBSP} to \ref{sec.MLDR} comprise the literature survey and description of various MOR methodologies classified into four main categories. The review concludes with a discussion of the presented methods and prospects for future research directions.

\section{Model order reduction}\label{sec.MOR}
The overall dynamics of a power system can be represented by a set of nonlinear time-invariant differential--algebraic equations (DAEs) of the form
\begin{align}
\dot{x} &= f(x,z,u,p), \label{eq.1}\\
0 &= g(x,z,u,p),\\
y &= h(x,z,u,p), \label{eq.2}
\end{align}
where $x\in\mathbb{R}^{n_x}$ denotes the vector of dynamic state variables, $z\in\mathbb{R}^{n_z}$ represents the algebraic variables associated with the network constraints, $u$ denotes external inputs or disturbances, $p$ is the vector of system parameters, and $y\in\mathbb{R}^{n_y}$ is the vector of output variables.

The aim of reduced-order modeling is to construct a lower-dimensional model of the form
\begin{align*}
\dot{x}_r &= f_r(x_r,z_r,u,p),\\
0 &= g_r(x_r,z_r,u,p),\\
y_r &= h_r(x_r,z_r,u,p),
\end{align*}
where $x_r\in\mathbb{R}^{n_{x,r}}$, $z_r\in\mathbb{R}^{n_{z,r}}$, and $y_r\in\mathbb{R}^{n_{y,r}}$ denote the reduced vectors of dynamic states, algebraic variables, and outputs, respectively, with $n_{x,r}<n_x$. The reduced model should accurately reproduce the input--output behavior of the original system while significantly reducing its computational complexity. More precisely, the approximation error should remain below a prescribed tolerance,
\begin{equation*}
|y-y_r|<\varepsilon,
\end{equation*}
where $\varepsilon>0$ is a user-defined error tolerance.

\subsection{Singular perturbation viewpoint for power-system DAE models}

A singular-perturbation (SP) interpretation of power-system dynamics has been widely applied in conventional SG-based power systems. There is a clear timescale separation of electromechanical dynamics of SGs and electromagnetic dynamics of transmission lines.

Let the vector of state variables is partitioned into slow and fast states as
\[
x =
\begin{pmatrix}
	x_s \\
	x_f
\end{pmatrix},
\qquad
x_s \in \mathbb{R}^{n_s}, \quad
x_f \in \mathbb{R}^{n_f}, \quad
n_s + n_f = n_x .
\]
After appropriate scaling of the
fast dynamics with a characteristic time constant, the differential–algebraic
power-system model can be written in the SP form
\begin{align}
	\dot{x}_s &= f_s(x_s,x_f,z,u,p), \label{eq:slow1}\\
	\varepsilon \dot{x}_f &= f_f(x_s,x_f,z,u,p), \label{eq:fast}\\
	0 &= g(x_s,x_f,z,u,p), \label{eq:alg}
\end{align}
where $0<\varepsilon \ll 1$ denotes the ratio between characteristic fast and
slow time scales. 

{\sf Reduced slow model.}
In the singular limit $\varepsilon\to 0$, the fast differential equation in~\eqref{eq:fast} is replaced by the algebraic constraint
\begin{equation*}
	0 = f_f(x_s,x_f,z,u,p),
\end{equation*}
which, together with the algebraic relation~\eqref{eq:alg}, forms a system of algebraic equations that locally
determines the values of the fast states and algebraic variables as functions of the slow states.
Geometrically, the set of all points satisfying these equations is called the
slow constraint manifold, which can be represented as
\begin{align*}
(x_f,z) = \Phi(x_s,u,p).
\end{align*}
Substituting this relation into the slow equation~\eqref{eq:slow1} leads to the
reduced-order model
\begin{equation*}
	\dot{x}_s
	=
	f_s\big(x_s,\Phi(x_s,u,p),u,p\big).
\end{equation*}
This reduced system describes the dominant long-term dynamics under the assumption that the fast variables instantaneously relax to their quasi-steady-state values determined by the slow variables. The fast state variables are therefore residualized, rather than discarded: instead of being neglected, they are represented implicitly through the algebraic mapping $\Phi(x_s,u,p)$. This distinguishes the quasi-steady-state approximation from truncation, in which the fast variables are simply omitted from the model.

\begin{figure}[!t]
	\centering
	\includegraphics[scale=0.7]{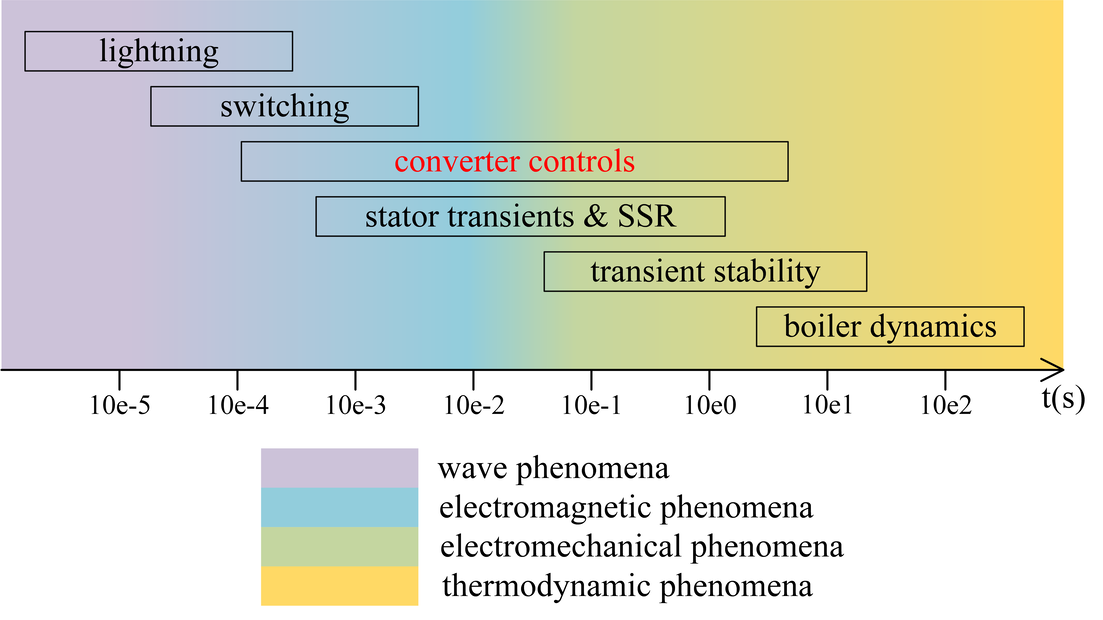}
	\caption{Power system dynamic phenomena and the associated time constants}
	\label{fig3}
\end{figure}

The main classes of dynamic phenomena in power systems and their associated time constants are illustrated in Figure \ref{fig3}. Traditional stability problems, such as transient and small-signal electromechanical stability, are typically analyzed using RMS-type models in which fast transmission grid dynamics is neglected (quasi steady-state or phasor-based approximation). In contrast, faster electromagnetic phenomena -- including switching transients and
lightning-induced disturbances -- are commonly investigated using electromagnetic transient (EMT) 
simulation tools, which rely on detailed component-level modeling and are 
therefore usually applied to smaller subsystems of interest.
 
\subsection{Multi-timescale nature of PED-based power systems}
\label{sec2.2}

The increasing deployment of VSC-based PEDs has significantly increased the complexity of the traditional classification of power system dynamic phenomena. Owing to the wide frequency bandwidth of converter control systems, the characteristic time constants of outer control loops are often comparable to those associated with electromechanical oscillations of SGs, while the time constants of ICC loops are of the same order as electromagnetic network modes. Consequently, converter-dominated power systems exhibit pronounced multi-timescale behavior, which substantially complicates the development of accurate and computationally efficient reduced-order models. The SP approximation cannot be straightforwardly applied.

Moreover, converter control architectures themselves possess an inherent hierarchical time-scale structure, with nested control loops operating over several orders of magnitude in time. This internal multi-timescale nature further contributes to the complexity of modeling, analysis, and stability assessment in modern power systems with high penetration of PE interfaces.

In summary, the derivation of structured reduced-order representations of converter-rich power systems remains a nontrivial task in the context of SP modeling. This difficulty stems primarily from two key aspects:

\begin{enumerate}
	\item[P1:] 
	Converter-dominated power systems exhibit pro\-noun\-ced multi-timescale dynamical behavior. Owing to the wide bandwidth of converter control systems and the interaction between electromagnetic and electromechanical processes, a clear separation of state variables based solely on time-scale considerations is often not possible.
	\item[P2:]
	Converter-driven stability phenomena may arise at substantially higher frequencies than those typically encountered in SG-dominated systems. Consequently, reduced-order models that retain only slow electromechanical dynamics risk neglecting critical high-frequency interactions associated with converter controls and network electromagnetic modes.
\end{enumerate}

\subsection{Overview of the literature}

\begin{figure}[!t]
	\centering
	\includegraphics[scale=0.5]{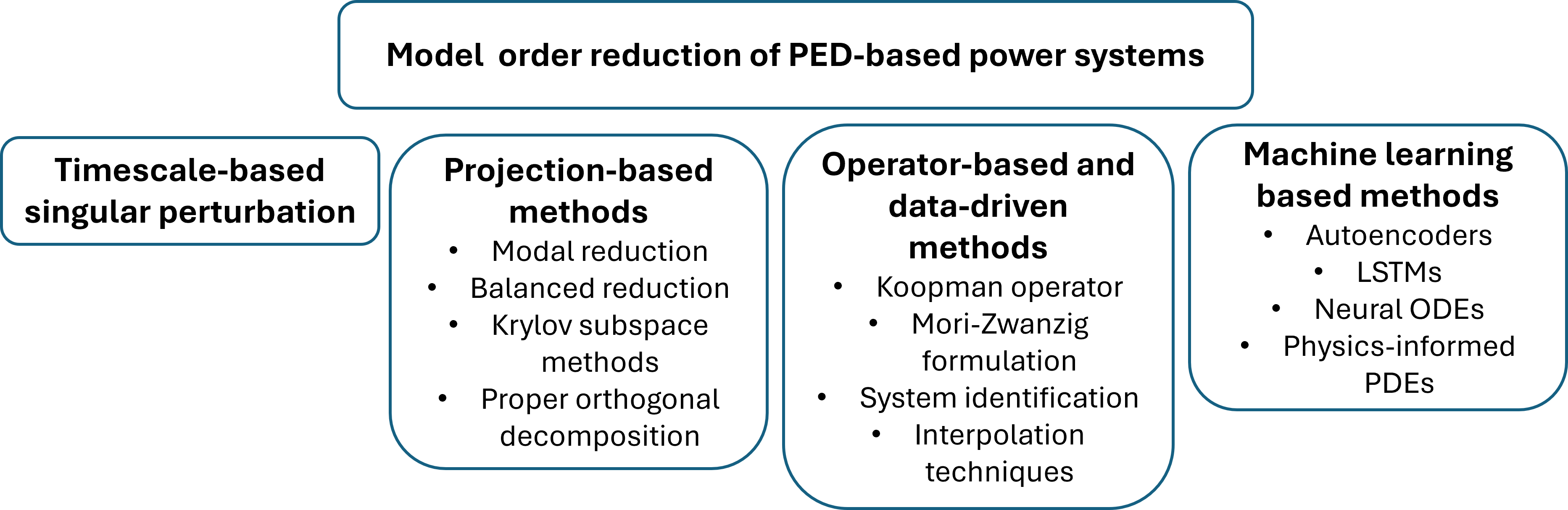}
	\caption{Proposed classification of MOR techniques}
	\label{fig:figure4}
\end{figure}

Based on the exhaustive literature review of different MOR techniques applied to PED-based power systems, the classification presented in Fig. \ref{fig:figure4} is introduced. The main categories include a timescale-based SP, projection-based methods, operator-based and data-driven methods, and machine learning (ML) algorithms. It needs to be emphasized that this classification is not unique, and there are also other ways to classify methodologies. Also, some of the proposed algorithms use a combination of different methods. The highest number of articles in the literature employs projection-based methods, especially the one founded on the modal analysis. The methods can generally be intrusive, which are applied to white-box models with known governing equations, and non-intrusive, which can be applied to both white-box and black-box models. The latter becomes especially relevant in the renewable energy landscape when system equations (e.g., control algorithms) are under IP protection, and only measurements are available. 

The review is mostly focused on reduced-order modeling of the whole power systems, including different elements, whether it is the case of large transmission-level systems, or distribution-level microgrids. However, examples of MOR techniques of individual power system elements, primarily wind farms and PV power plants, are also included when there is a recognized potential for their application to the system-level reduction. A few MOR examples from SG-based power systems, interesting from the perspective of the PED-based power systems, are also included in this review.

\section{Timescale-based singular perturbation}\label{sec.TBSP}

The most direct approach to constructing a SP model, \eqref{eq:slow1}-\eqref{eq:alg}, is based on the classical small-parameter methodology traditionally applied in SG-dominated power systems. This approach is also employed in networks with grid-forming converters \cite{Ajala2023}, where grid and fast converter dynamics are eliminated by exploiting their relatively short time constants. In such procedures, the differential equations are ordered according to their characteristic time scales, and those associated with sufficiently small time constants are replaced by algebraic constraints, leading to a reduced slow subsystem. A timescale-based SP is also applied in \cite{11141565} for the purposes of the large-disturbance stability evaluation for a power system with multiple virtual synchronous generator grid-forming converters. However, the simplified converter models are employed, not including inner control loops, the reactive power outer control loop, and filter dynamics.

Ma et al.~\cite{MaWangYuanHong2024} applied a timescale-based SP method to the reduction of converter-based microgrids. In addition to deriving the reduced-order model, they performed a rigorous analysis of its accuracy and stability by introducing the boundary layer model (BLM), which captures the fast dynamics. The authors proved that if the BLM is globally asymptotically stable and the reduced-order model is input-to-state stable, then the original full-order model is also input-to-state stable. Here, global asymptotic stability means that every trajectory converges to the equilibrium point regardless of the initial condition, whereas input-to-state stability guarantees that bounded inputs produce bounded state trajectories.

Instead of applying SP to state variables, ref. \cite{XuCheng2024PESGM10688582} applies it to the state matrices. Cosine similarity is used to determine transition points between different options of reduced-order models. 
However, the method was only applied to the grid-forming converter model.

\section{Projection based methods}

The main projection-based techniques for developing reduced-order models include modal reduction, balanced reduction, Krylov subspace methods, and proper orthogonal decomposition (POD). The first three approaches are based on linearized system models, whereas POD can be applied directly to nonlinear systems without prior linearization. Among these methods, modal reduction is by far the most widely adopted in the literature.

\subsection{Modal reduction}

To address the aforementioned challenges in the construction of SP models, researchers have frequently relied on linear small-signal analysis techniques. A prerequisite for such analyses is the linearization of the nonlinear power-system model, Eqs. \eqref{eq.1}-\eqref{eq.2}, around a given steady-state operating point. This leads to the state-space representation
\begin{align}
	\Delta \dot{x} &= A \Delta x + B \Delta u, \label{eq:lin1}\\
	\Delta y &= C \Delta x + D \Delta u, \label{eq:lin2}
\end{align}
where $A$ denotes the state (system) matrix, $B$ the input (control) matrix, $C$ the output (measurement) matrix, and $D$ the feedthrough matrix.

The inherent dynamic characteristics of the system are obtained from the eigenvalues $\lambda = \sigma+j\omega$ of the state matrix $A$. Small-signal stability requires that the real parts of all system eigenvalues are negative.

\subsubsection{Participation factor analysis}

Participation factor analysis (PFA) is one of the most widely used modal-analysis techniques for identifying the state variables that contribute most significantly to individual dynamic modes. It is particularly valuable for large-scale power-system models, where the high dimensionality of the state space makes it difficult to determine the sources of poorly damped oscillations and small-signal instability.

The participation factor of the $k$-th state variable in the $i$-th mode is defined as
\begin{equation*}
p_{ki}=u_{ik}v_{ki},
\end{equation*}
where $v_{ki}$ and $u_{ik}$ denote the $k$-th components of the right and left eigenvectors, respectively. Participation factors quantify the degree to which a state variable contributes to a particular dynamic mode. A large value of $p_{ki}$ indicates that the $k$-th state variable has a strong influence on the mode associated with $\lambda_i$, whereas small values indicate only a minor contribution.

Based on modal analysis and PF-based state classification, several studies have proposed structured approaches for reduced-order modeling of converter-dominated systems. For example, in the analysis of grid-forming hybrid renewable-energy microgrid clusters~\cite{Zhao2024}, the overall state vector is decomposed into three groups: global slow dynamics, coupling dynamics, and global fast dynamics.  Reduced-order models are then constructed by retaining the coupling states together with the dominant fast dynamics, while less influential states are aggregated or eliminated.

Zhao et al.~\cite{ZHAO2024110331} employed modal analysis together with PFA to classify the state variables of a hybrid wind--solar microgrid into slow and fast subsets. The resulting reduced-order model enabled computationally efficient reachability analysis under uncertainty.

\subsubsection{Iterative construction of SP models}
In converter-dominated power systems, the presence of several interacting
time scales often prevents a direct decomposition of the system dynamics
into clearly separated slow and fast subsystems. In such cases, the
construction of SP models is typically performed
using iterative modal-based reduction procedures \cite{Nayak2021}.

In the initial step, PFA is employed to identify participating states associated with the
fastest dynamics which are then aggregated or eliminated, yielding a reduced-order
model of the form
\begin{equation*}
	\Delta \dot{x}^{(1)} = A^{(1)} \Delta x^{(1)}.
\end{equation*}

Since the reduction step alters the spectral properties of the system,
eigenvalue analysis is repeated for the reduced matrix $A^{(1)}$. If the
resulting model still exhibits multiple interacting time scales, the
clustering and reduction procedure is applied again. Through successive
iterations, the system dynamics are progressively transformed into a
two-time-scale structure that can be represented in the SP form
\begin{align}
	\Delta \dot{x}_s &= A_{ss}\Delta x_s + A_{sf}\Delta x_f, \label{eq:sp_slow}\\
	\varepsilon \Delta \dot{x}_f &= A_{fs}\Delta x_s + A_{ff}\Delta x_f.
	\label{eq:sp_fast}
\end{align}

\subsubsection{Participation-based grouping and adaptive residualization}

In large-scale power-system models, individual dynamic modes are typically
associated with the collective behavior of several state variables rather
than with a single state. Conversely, a given state variable may participate
in multiple oscillatory modes. For this reason, modal-based model-reduction
approaches often rely on the construction of groups of states and the
corresponding groups of eigenvalues.

A PF-based MOR methodology proposed in~\cite{Cossart2022} constructs reduced-order representations by
systematically identifying groups of state variables associated with
individual eigenmodes. For a given eigenvalue $\lambda_i$, a set of dominant state variables
$\mathcal{P}_{\lambda_i,\varepsilon_{\mathrm{part}}}$ is first formed
so that the states corresponding to the largest PFs
are retained until their cumulative contribution exceeds a prescribed
threshold $\varepsilon_{\mathrm{part}}$. Formally, the set is defined as
\begin{equation*}
	\mathcal{P}_{\lambda_i,\varepsilon_{\mathrm{part}}}
	=
	\left\{
	x_k \;\middle|\;
	p_{ki} \ge p_{mi} \ \forall x_m \notin \mathcal{P}_{\lambda_i,\varepsilon_{\mathrm{part}}},
	\;
	\sum_{x_k \in \mathcal{P}_{\lambda_i,\varepsilon_{\mathrm{part}}}} p_{ki}
	\ge \varepsilon_{\mathrm{part}}
	\right\}.
\end{equation*}

Based on these sets, groups of eigenvalues and the corresponding groups
of state variables are established. Residualization is then performed
at the level of entire state groups. Eliminating all state variables
belonging to a given group results in the removal of the associated
eigenvalues while producing only limited shifts in the remaining system
spectrum. This property enables structured reduction of large dynamical
models while preserving dominant modal characteristics.

A related AC grid MOR methodology presented in~\cite{Grdenic2022}
is based on the identification of subsystem interaction modes. These are
defined as eigenmodes exhibiting significant participation of state
variables associated with both a PED and a
transmission-line model. For each transmission line, the set of interaction
modes is determined as
\begin{equation*}
	\mathcal{I}
	=
	\left\{
	\lambda_i \;\middle|\;
	\sum_{x_k \in \mathcal{A}_{\mathrm{line}}} p_{ki} > \eta_{\mathrm{line}},
	\;
	\sum_{x_k \in \mathcal{A}_{\mathrm{PED}}} p_{ki} > \eta_{\mathrm{PED}}
	\right\},
\end{equation*}
where $\mathcal{A}_{\mathrm{line}}$ and $\mathcal{A}_{\mathrm{PED}}$
denote the sets of state variables describing the transmission line and
the PED subsystem, respectively, while $\eta_{\mathrm{line}}$ and
$\eta_{\mathrm{PED}}$ are prescribed participation thresholds.

The identification of such interaction modes provides guidance for
selective reduction of transmission-line model complexity in the reduced
system representation. The choice of subsystem participation thresholds
determines the trade-off between computational efficiency and modeling
accuracy.

\subsubsection{Model-stiffness-oriented reduction and modal decoupling}

An alternative MOR strategy proposed in~\cite{Muntwiler2024}
is aimed at reducing the numerical stiffness of transmission-system models
with a high penetration of inverter-based resources. Model stiffness is
characterized by the ratio between the largest and the smallest real parts
of the system eigenvalues, defined as
\begin{equation*}
	\rho
	=
	\frac{
		\max\limits_{\lambda_i \in \sigma(A)} \, \Re(\lambda_i)
	}{
		\min\limits_{\lambda_i \in \sigma(A)} \, \Re(\lambda_i)
	},
\end{equation*}
where $\sigma(A)$ denotes the spectrum of the state matrix $A$. Large stiffness ratios indicate the coexistence of very fast and very slow dynamic modes, which significantly increases the computational burden of
time-domain simulations.

The reduction methodology in~\cite{Muntwiler2024} relies on a modal decoupling of
the linearized system dynamics. By transforming the state matrix into its
Jordan canonical form, the dynamic states can be represented in terms of
independent modal coordinates. In this representation, the participation
factor matrix assumes a block-diagonal structure, reflecting the complete
decoupling of individual modal subsystems. Based on this modal decomposition, a SP reduced-order
model is constructed by retaining a prescribed number of the slowest
eigenmodes, while the faster modes are neglected. This procedure reduces
the stiffness of the resulting dynamical model and improves computational
efficiency.

\subsubsection{Influence of fast dynamics: zero- and first-order singular perturbation models}

A generalization of SP-based model reduction that accounts for the influence of fast dynamics on the slow subsystem was proposed in~\cite{Vorobev2018} and demonstrated for inverter-based droop-controlled AC microgrids.

{\sf Zero-order singular perturbation (ZOSP).}
The classical SP approximation assumes that the fast subsystem instantaneously reaches its quasi-steady state by setting
$
\Delta\dot{x}_f=0
$ in Eq. \eqref{eq:sp_fast}.
This yields
\begin{equation*}
\Delta x_f=-A_{ff}^{-1}A_{fs}\Delta x_s,
\end{equation*}
which, when substituted into~\eqref{eq:sp_slow}, gives the reduced slow subsystem
\begin{equation}
\Delta\dot{x}_s =
\left(A_{ss}-A_{sf}A_{ff}^{-1}A_{fs}\right)\Delta x_s= A_r\Delta x_s.
\label{eq:zosp}
\end{equation}

The stability of the reduced-order model is determined by the eigenvalues of the corresponding reduced system matrix $A_r$. Equivalently, asymptotic stability can be verified using the classical Routh--Hurwitz criterion, which provides necessary and sufficient algebraic conditions on the coefficients of the characteristic polynomial of $A_r$.

{\sf First-order singular perturbation (FOSP).}
To account for the dynamic influence of the fast subsystem, the ZOSP can be refined by retaining the first-order term in the SP expansion while neglecting higher-order terms. 
This yields the first-order SP model
\begin{equation}
\Delta \dot{x}_s =
\left(I+\varepsilon A_{sf}A_{ff}^{-2}A_{fs}\right)^{-1} A_r\Delta x_s.
\label{eq:fosp}
\end{equation}
The correction term in \eqref{eq:fosp} captures the influence of the fast dynamics on the slow subsystem.

Consequently, even when the ZOSP~\eqref{eq:zosp} predicts a stable reduced-order model, stability of the first-order model is not guaranteed. The additional coupling introduced by the fast dynamics may alter the stability properties of the reduced system. This effect is particularly important in inverter-based microgrids, where transmission lines often exhibit a relatively high resistance-to-reactance ratio, resulting in stronger coupling between the network electromagnetic dynamics and the converter controls.

The FOSP methodology was also employed in \cite{9736580} to develop a reduced-order model of an islanded microgrid with grid-forming converters.

\subsubsection{Three-time-scale SP reduction}

An extension of the previously described SP
reduction strategies is presented in~\cite{Caduff2021}, aiming to combine
the low computational complexity of the ZOSP model with the improved accuracy of the FOSP approximation. The key idea is to represent the system dynamics using three distinct time scales. In the resulting reduced-order modeling framework, the dynamics of the
slow subsystem are retained explicitly. The fast subsystem is approximated
using the FOSP approach, whereby the first
derivatives of the fast states are preserved while higher-order derivatives
are neglected. In contrast, the dynamics of the very fast states are
completely neglected according to the ZOSP approximation, resulting in algebraic constraints for these variables.

The classification of system states into the three disjoint groups is
performed using an iterative PFA. In this procedure, participation factors are recomputed after each
residualization step, allowing a progressive refinement of the
time-scale separation and identification of dynamically dominant
state variables.

\subsubsection{Component connection methods}

The MOR approach proposed in \cite{DewanganPuricelliBeerten2026} is focused only on the preservation of the interaction modes in the interconnected system and the corresponding state variables participating in those modes. To this end, the authors employ the component connection method (CCM) to assemble the complete system model and quantify the interactions between individual subsystems. The state-space matrix of the interconnected system is expressed as
 \begin{equation*}
	A = A_d + A_c, \qquad
	A_c = B_d L_1 
	\left( I - D_d L_1 \right)^{-1} C_d
\end{equation*}
Here, $A_d$ is the block-diagonally augmented state-space matrix 
of the subsystem state-space matrices, and $A_c$ is the coupling matrix. 
Matrices $B_d$, $C_d$, and $D_d$ are the 
block-diagonally augmented input, output, and feedthrough matrices of the 
subsystems, respectively 
\begin{equation*}
X_d=
\operatorname{diag}(X_1,X_2,\ldots,X_N),
\qquad
X\in{A,B,C,D},
\end{equation*}
where $X_d$
denotes the block-diagonal matrix constructed from the corresponding subsystem matrices $X_i$. Matrix $L_1$ is the Jacobian of the subsystem input vector with respect to the subsystem output vector.

To reduce the computation time and improve tractability, the interaction modes are determined from the open-loop eigenvalues of the individual subsystems. First, the open-loop eigenvalues and eigenvectors of the individual subsystems are calculated. Then, based on the eigenvalue proximity criterion and eigenvalue displacement criterion, the interaction modes are determined. Once the interaction modes are established, the state variables with low participation are residualized in the individual subsystems. Finally, the CCM is used to assemble the reduced-order model of the entire system.

\subsubsection{Residue-based modal reduction}

If the system from Eqs. \eqref{eq:lin1}-\eqref{eq:lin2} is transformed into the Laplace domain, the transfer function can be written as:
\begin{equation}
	G(s) = \sum_{i=1}^{n} \frac{R_i}{s - \lambda_i} + D \label{trf}
\end{equation}
if there are no repeating eigenvalues. Here, $R_i$ is the residue corresponding to the eigenvalue $\lambda_i$, and it can be calculated as:
\begin{equation*}
	R_i = C \, u_i \, v_i^\mathsf{T} B.
\end{equation*}

Using this approach, Zhao et al. \cite{ZhaoWuLuoXieYangNiLai2024} developed a reduced-order model for interconnected microgrids incorporating wind and PV power plants. First, the full system is partitioned into slow and fast subsystems according to the characteristic time constants of the state variables. Next, the dominant modes of the fast subsystem are identified using the residue-based indicator
\begin{equation} \label{eq:residue}
\bar{R}_i=
\frac{|R_i|_2}{|\operatorname{Re}(\lambda_i)|},
\end{equation}
which combines the modal residue, reflecting the controlability and observability of the mode, with its decay rate through the real part of the corresponding eigenvalue. Finally, PFA is employed to identify the fast state variables that contribute most significantly to the selected modes. These state variables are retained in the reduced-order model, while the remaining fast variables are eliminated.

Residues defined by Eq. \eqref{eq:residue} and PFA are used in \cite{10669059} for developing a reduced-order model of the grid-tied hybrid system consisting of a PV, wind turbine and battery energy storage system. The Subspace Accelerated MIMO Dominant Pole algorithm is applied in \cite{8425793} to extract only the dominant modes of large wind farm models, and thus, shorten the calculation procedure. This algorithm includes an iterative method to find a pole with the largest indicator as defined by Eq. \eqref{eq:residue}.

\subsubsection{Segmented linearization}

The reduced-order models based on linearization are the most accurate near the linearization points, but become less accurate as we move farther away. To overcome this drawback, articles \cite{10359155} and \cite{electronics15010157} conduct the so called segmented linearization. Multiple linearized systems are calculated for a changing operating point during the fault scenario for the permanent magnet synchronous generator (PMSG) based wind farm. To distinguish the importance of the states, as a prerequisite for the subsequent MOR, the dominance degree and the generalized participation factor measures are introduced. The dominance degree measures the influence of a certain eigenvalue on the system output:
\begin{equation*}
	T_{i\text{-}jh} = \frac{b_{\Lambda ih}\, c_{\Lambda ji}}{\lambda_i}
\end{equation*}
where $b_{\Lambda ih}$ and $c_{\Lambda ji}$ are the corresponding modal controlability and observability projections. The generalized participation factor combines the classical participation factor and the dominance degree:
\begin{equation*}
	w^{s}_{ki} = u^{s}_{ki} \, v^{s}_{ki} \, s^{s}_{i}
\end{equation*}
with $s^{s}_{i}$ being the overall dominance of mode $\lambda_i$.

\subsubsection{Other modal-based MOR techniques}

Yan et al. \cite{YAN2024109835} applied transfer-based MOR by preserving the open-loop characteristics of the reduced-order model of a microgrid with voltage source inverters. Zeros and poles associated with fast dynamics are eliminated, while their influence is preserved via the open-loop gain.

The paper \cite{11119160} leverages the fact that the eigenvalues of the large-scale converter multi-infeed-system can be approximated by the eigenvalues of the smaller double-infeed-subsystems. This is achieved by creating an appropriate transformation matrix $T$ to block-diagonalize the Kron-reduced bus admittance matrix $B_{S}$ by minimizing the magnitude of off-diagonal terms:
\begin{equation*}
	TB_ST^{-1}=\Lambda_D
\end{equation*}
The proposed methodology is demonstrated on several test cases, including converters in the grid-following and the grid-forming operation modes.

\subsection{Balanced reduction}

In addition to modal-analysis-based approaches, balanced realization
constitutes another widely used linear analysis technique in MOR methodologies. This framework evaluates the
dynamic importance of system states based on their controlability
and observability properties. For an asymptotically stable system, the controlability
and observability Gramians are defined as
\begin{align*}
	W_{\mathrm{c}}
	&=
	\int_{0}^{\infty}
	e^{A t} B B^{\top} e^{A^{\top} t} \, dt, \\
	W_{\mathrm{o}}
	&=
	\int_{0}^{\infty}
	e^{A^{\top} t} C^{\top} C e^{A t} \, dt .
\end{align*}
Equivalently, these matrices are obtained by solving the Lyapunov equations
\begin{align}
AW_{\mathrm{c}}
+W_{\mathrm{c}}A^{\top}
+BB^{\top}
&=0,\label{lyapunov1}\\
A^{\top}W_{\mathrm{o}}
+W_{\mathrm{o}}A
+C^{\top}C
&=0.\label{lyapunov2}
\end{align}
The controlability Gramian $W_{\mathrm{c}}$ quantifies how effectively each state can be influenced by the system inputs, whereas the observability Gramian $W_{\mathrm{o}}$ measures how strongly each state affects the system outputs. For a minimal stable system, an appropriate similarity transformation yields a balanced realization in which the two Gramians are equal and diagonal,
\begin{equation*}
W_{\mathrm{c}} =
 W_{\mathrm{o}} =
\Sigma,
\end{equation*}
where
\[
\Sigma=\operatorname{diag}(\sigma_1,\ldots,\sigma_n)
\]
contains the Hankel singular values. These values quantify the joint controlability and observability of the system states. 
States associated with small Hankel singular values contribute little to the input–output dynamics and can therefore 
be truncated or residualized with only a limited effect on the system response.

A limitation of balanced reduction in power-system applications is that the balancing transformation generally destroys the direct physical interpretation of the state variables. Consequently, the method is not always suitable when structured model reduction or preservation of subsystem dynamics is required. Nevertheless, balanced reduction offers a rigorous a priori error bound on the reduced-order model,
\begin{equation*}
\|G(s)-G_r(s)\|_{\infty}
\le
2\sum_{i=r+1}^{n}\sigma_i,
\end{equation*}
where $\sigma_i$ are the discarded Hankel singular values. 
This bound guarantees that the approximation error remains small provided that the neglected Hankel singular values decay sufficiently rapidly.

\subsubsection{Combining Hankel singular values and participation factors}

In \cite{TomasMartin2023}, the relevance of
microgrid state variables is quantified by combining Gramian-based measures with
modal-analysis indicators. The resulting relevance coefficients are
then used to guide the construction of reduced-order models that balance
computational efficiency and dynamic fidelity. The authors in \cite{avendano2026application} also used state relevance coefficients (combining Hankel singular values and participation factors) for the MOR of a virtual power plant consisting of several converter-interfaced wind and PV power plants. Moreover, it is shown that a smaller number of states is required with balanced truncation; however, the physical interpretability of states, as explained, is not preserved in this case. The authors in \cite{11234583} combined PFs for poorly damped modes with Hankel singular values to determine which states will be residualized in the full-order PMSG-based wind farm model in the grid-forming mode.

\subsubsection{Frequency-limited Gramians}

The concept of frequency-limited controlability and observability Gramians is used in \cite{deMoraisDiasCamposSarnetKilter2023} to develop reduced-order models in a desired frequency range:
\begin{align*}
	W_{C,\omega} &= \int_{\omega_1}^{\omega_2} 
	\left( j \omega I - A \right)^{-1} 
	B B^* 
	\left( -j \omega I - A^* \right)^{-1} 
	\, d\omega, \\
	W_{O,\omega} &= \int_{\omega_1}^{\omega_2} 
	\left( - j \omega I - A^* \right)^{-1} 
	C^* C 
	\left( j \omega I - A \right)^{-1} 
	\, d\omega.
\end{align*} 
The participation of the state $k$ in the input-output behavior over the interval $[\omega_1, \omega_2]$ is defined as:
\begin{equation}
	p_\omega(k) =
	\frac{W_{O,\omega}(k,k) \, W_{C,\omega}(k,k)}
	{\sum_{i=1}^{n} W_{O,\omega}(i,i) \, W_{C,\omega}(i,i)}
	\label{eq:pomega}
\end{equation}
The states with participation higher than a predefined threshold are retained in the reduced-order model, and the other states are residualized. The methodology is demonstrated on the modified IEEE 39-bus grid with two converters.

Similarly, the frequency-weighted balanced residualization is conducted in \cite{9051711} by introducing input and output weighting transfer functions. The methodology is successfully tested on two power system models also including Pad\'e approximations for time delays of the wide-area controller; however, converters are not included in the test systems. The frequency-weighted and frequency limited methods are also applied in \cite{10489949} for the MOR of doubly fed induction generator (DFIG) based wind power plants.

\subsubsection{Time-weighted Gramians}

Time-weighted Gramians are used in \cite{10921669} for reducing the model order of a DFIG-based wind power plant, which represents a time-sensitive system:	
\begin{align*}
		W_{Cw} = \int_{0}^{\infty}w(t) 
		e^{A t} B B^\top 
		e^{A^\top t} \, dt \\
		W_{Ow} = \int_{0}^{\infty}w(t) 
		e^{A^\top t} C^\top C 
		e^{A t} \, dt
\end{align*}
The weighting functions $w(t)$ in the above equations enable emphasis on system behavior in the specified time interval (e.g., transient response), compared to classical Gramians, which capture system behavior over an infinite time horizon.

\subsubsection{Cross-Gramian}

Instead of using controlability and observability Gramians, the paper \cite{8744409} employs the cross-Gramian $G_{co}$, which can be obtained as a solution of the Sylvester equation:
\begin{equation*}
	A G_{co} + G_{co} A + BC = 0 
\end{equation*}
The cross-Gramian combines system controlability and observability into a single measure without the need to calculate both Gramians. The methodology is demonstrated on classical SG-based power systems with poorly damped inter-area oscillation modes, in which the external subsystem is reduced, and the SG of interest (with the power system stabilizer) remains unaltered.

\subsubsection{Specialized balanced truncation methods}

Wang et al. in \cite{9319514} applied balanced truncation to a dc microgrid with inhomogeneous initial conditions, e.g., during the soft start of converters when the model state vector deviates from the equilibrium point. The model output using the Duhamel equation can be decomposed into two parts:
\begin{equation*}
	y(t)
	=
	Ce^{At}x_0
	+
	\int_0^t
	Ce^{A(t-\tau)}
	Bu(\tau)\, d\tau
\end{equation*}
The first part corresponds to the unforced response due to nonzero initial conditions $x_0$, and the second part to the forced response due to the input $u(t)$. The balanced truncation is applied separately to those two parts, and the resulting reduced model is obtained by aggregating them.

The methodology in \cite{YuWangWangPanSun2024} is focused on preserving the PLL dynamics. The system model, consisting of 20 DFIG wind turbines, is divided into two subsystems: decoupled subsystem containing PLL dynamics and the main system (the rest of the system). The proposed method preserves PLL dynamics while the rest of the system is reduced using balanced truncation.

\subsection{Krylov subspace methods}

Krylov subspace methods, also known as moment-matching methods, constitute another important class of projection-based MOR techniques. Their objective is to construct a reduced-order model whose transfer function matches the behavior of the original system around a prescribed expansion point $s_0\in\mathbb{C}$.
The transfer function from Eq.~\eqref{trf} can be expanded into a Taylor series around $s_0$,
\begin{equation*}
G(s) =
\sum_{n=0}^{\infty}
M_n(s-s_0)^n,
\end{equation*}
where $M_n$ are referred to as the moments of the transfer function. The aim is to determine a reduced transfer function $G_r(s)$ 
whose first $k$ moments coincide with those of the original system. The most widely used technique for constructing the Krylov subspace and computing the moments is the Arnoldi method. 
It generates an orthonormal basis of the Krylov subspace
\begin{equation}
	\mathcal{K}_r(A, B) 
	= \text{span}\{B, A B, A^2 B, \dots, A^{r-1} B\}
	\label{eq:krylov_subspace}
\end{equation}
which is subsequently used to project the original system onto a lower-dimensional subspace. The resulting reduced-order model is obtained by means of a Galerkin projection and preserves the prescribed moments of the original transfer function.

The Krylov methods have been used in developing reduced-order models of SG-based power systems, e.g. \cite{1425586}, but are lately being applied to modern renewable-based power systems as well. The authors in \cite{AsgarimoghaddamWang25} compared balanced reduction and the Krylov subspace method in developing a reduced-order model for the IEEE 37 bus grid with inverter-based generation. The obtained results confirm the theoretical premises: balanced reduction is superior for accuracy over a broad frequency range and the stability preservation of the reduced model. The Krylov subspace method is more computationally efficient. However, a credible approximation is limited to a narrow frequency band, and the method does not guarantee the preservation of stability properties.

\subsubsection{Iterative Rational Krylov Algorithm}

The Iterative Rational Krylov Algorithm (IRKA) is employed in \cite{AliPal2021} in developing a reduced-order model of two large offshore wind farms connected to a multi-terminal grid. The goal of the algorithm is to minimize the error:
\begin{equation*}
	\min_{G_r \in \mathcal{H}_2^{(r)}} 
	\left\| G(s) - G_r(s) \right\|_{\mathcal{H}_2}
\end{equation*}
The rational Krylov subspace is defined as:
\begin{equation*}
	\mathcal{K}_r(A,B,\{\sigma_i\}_{i=1}^r)
	=
	\mathrm{span}\left\{
	(A - \sigma_i I)^{-1} B
	\right\}_{i=1}^r .
\end{equation*}
where $\sigma_i$ are the interpolation points (or shifts) in the complex plane. During the iterative procedure, these interpolation points are updated until convergence is achieved, yielding a reduced-order model that satisfies the first-order optimality conditions for $\mathcal{H}_2$-optimal approximation. The authors in \cite{8425793} showed superior properties of IRKA in reduced-order modeling of large wind power plants. To ensure preservation of the damping ratio and frequency of poorly damped modes in the IRKA method, the paper \cite{8239650} utilizes heuristic optimization to solve a multiobjective optimization problem. Unlike the Arnoldi algorithm, which employs fixed expansion points, IRKA iteratively updates the interpolation points, resulting in significantly improved approximation accuracy.

Similar to the approach presented in \cite{10359155}, the methodology proposed in \cite{8693742} approximates the nonlinear system by linearizing it at multiple operating points along the trajectory of the simulated disturbance. This trajectory piecewise-linear approach represents the full nonlinear model as a weighted combination of local linear models,
\begin{align*}
\dot{x} &=
\sum_{i=0}^{s-1}
\alpha_i(x,x_i)
\left[
f(x_i,u_i) + A_i(x-x_i) + B_i(u-u_i)
\right],\\
\end{align*}
where the weighting coefficients $\alpha_i(x,x_i)$ depend on the distance between the current system state and the corresponding linearization points. Each local linear model is subsequently reduced using IRKA, although other projection-based MOR techniques can also be employed.

\subsubsection{Extended Krylov subspace method}
A valuable work on reducing the model order of a complex power system model is conducted in \cite{Du2024ParallelPowerSystem}, although it does not include PEDs. An extended Krylov subspace has been used for improving computational efficiency of solving Lyapunov equations. Beforehand, the $\alpha$ shift is applied to the system state-space matrix to enable the application of balanced realization to unstable systems. The proposed methodology is successfully applied to a 39-bus and 145-bus grids to obtain a reduced-order model for tuning a SG power system stabilizer.

\subsubsection{Parametric Krylov method}
The authors in \cite{10770089} employed the parametric Krylov method for model reduction, accounting for the model parameter variations (e.g., operating point, or controller coefficients).
The full model transfer function, in this case, can be expanded into a Taylor series around $s_0 \in \mathbb{C}$:
\begin{equation*}
	G(s, \lambda)
	=
	\sum_{n=0}^{\infty}
	M_n(\lambda) \,(s - s_0)^n
\end{equation*}
The reduced-order model does need to match not only across the frequency range, but also for the multiple parametric values. In the derivation of the reduced-order model, the frequency-domain adaptive method is employed, considering not only the convergence error but also the cumulative error and the rank of the projection matrix. The methodology is applied to converter-based power systems of different sizes. The stability is accurately assessed across a wide frequency range.

\subsubsection{Stability-preserving Krylov method}
A stability-preserving Krylov subspace projection algorithm is employed in \cite{ASGARIMOGHADDAM2026112546} to ensure stability of the reduced-order microgrid model. Unlike the conventional Krylov method using Galerkin projection in which left and right projection matrices are the same ($W=U$), the stability-preserving algorithm uses oblique projection ($W\neq U$) in which the right projection matrix ensures accuracy by matching moments, and the left projection matrix satisfies the Lyapunov stability condition.

\subsubsection{Structure-preserving Krylov method}
Wang et al. \cite{8859394} employed the Structure-Preserving Reduced-order Interconnect Macro-modeling (SPRIM) algorithm for the MOR of an active distribution network, which is modeled using the dynamic phasor framework. However, the reduction involves only passive RLC elements (branches, transformers, static loads, etc.), while other components (inverters, generators, etc.) were accounted for by boundary conditions. By using the SPRIM algorithm, the reduced-order model preserves a structure that is typical for RLC circuits.  

\subsubsection{Two-sided moment matching}
A signal generator is applied in \cite{10804098} to excite the full-order power system model at the interpolation points. A steady-state system response produces a data-driven transfer function representation at the selected frequencies. The two-sided moment matching is achieved by applying interpolation conditions on both the input and the output side. This way, a higher level of accuracy is achieved for the same number of interpolation frequencies. In the final step, the reduced-order model is constructed by solving a Sylvester equation where the size of the reduced-order model corresponds to the number of interpolation points. The methodology is tested for reducing a large wind farm model connected to the IEEE 14-bus system.

\subsection{Proper orthogonal decomposition}

The proper orthogonal decomposition (POD) is used much less in developing power system reduced-order models. Differently from the above described projection-based methods, a system linearization is not a prerequisite. First, the system needs to be simulated for a specific event to obtain a snapshot matrix:
\begin{equation*}
	X =
	\begin{bmatrix}
		x_1(t_1) & \cdots & x_{1}(t_m) \\
		\vdots & \ddots & \vdots \\
		x_{n}(t_1) & \cdots & x_{n}(t_m)
	\end{bmatrix}
\end{equation*}
The number of time steps $m$ needs to be much larger than the number of state variables $n$ to capture all relevant dynamics properly. Afterward, the eigenvalue decomposition of $X X^{T}$ is performed:
\begin{equation*}
	X X^T = U \Lambda U^T,
\end{equation*}
The matrix $\Lambda$ contains eigenvalues that are real and non-negative. The final step includes keeping the $r$ state variables that correspond to the $r$ largest eigenvalues, and other state variables are truncated or residualized from the system model.

The structure-preserving reduction methodology of the original nonlinear DAE power system model is presented in \cite{10753627} and \cite{11107785}, in which POD was separately conducted for the dynamic and for the algebraic variables. The authors also introduced a balanced POD by approximating controlability and observability matrices from the input-output response snapshots, as was done in \cite{7464373}, where it showed higher accuracy compared to balanced truncation. Both methods in \cite{10753627} are tested on IEEE 39-bus and 2000-bus systems, including PV power plants in grid-forming mode.

\section{Operator-based and data-driven methods}

This section describes operator-based and data-driven methods which are building reduced-order models based on the operator approximations or the observed system behavior.

\subsection{Koopman operator}
The article \cite{10838254} leverages the use of the Koopman operator to reduce large-scale power systems, with a focus on transient stability:
\begin{equation*}
	Kh(x)=h(F(x))
\end{equation*}
where $K$ is the Koopman operator, $h(x)$ are the observable functions of the system states, and $F(x)$ is the nonlinear flow map of the system. The Koopman operator is calculated from the simulation data of the system, transforming nonlinear system into a linear one in the space of observables. Finally, the reduction is conducted by eigenvalue decomposition of $K$ and retaining only the most relevant modes.

\subsection{Mori-Zwanzig formulation with Dynamic Mode Decomposition}
The authors in \cite{11197966} employed the Mori-Zwanzig (MZ) formulation which transforms nonlinear dynamics into a reduced linear representation with additional memory effect:
\begin{equation*}
	\frac{d\bar{x}(t)}{dt}
	=
	M\bar{x}(t)
	-
	\int_0^t K(t-s)\bar{x}(s)\,ds
	+
	F(t)
\end{equation*}
where $\bar{x}(t)$ denotes the preserved reduced-order system states, the first term represents their dynamics, the second term accounts for the impact of the eliminated states, and the third term represents the noise effects. In the second step of the methodology, Dynamic Mode Decomposition (DMD) is applied to the simulation data to obtain:
\begin{equation*}
	x(t)=\Phi e^{\Omega t}b
\end{equation*}
where $\Phi$ is the matrix of dynamic modes, $\Omega$ is the diagonal matrix of modal eigenvalues, and $b$ contains 
the corresponding modal amplitudes determined from the initial condition. 
By combining the MZ formulation with DMD, the proposed methodology 
enables parallel-in-time simulation of converter-dominated microgrids, resulting in a significant reduction in transient simulation time.

\subsection{Frequency-domain system identification}
In \cite{10310269}, reduced-order models of converters and cables are developed using the vector/matrix fitting algorithm from the impedance/admittance measurements. Subsequently, the whole system model is incrementally assembled using the proposed state-space series and parallel operators. 
In \cite{11134811}, first, the high-order combined transfer function is obtained, which includes impedances/admittances of converter-interfaced resources, lines, loads, etc. Later, the order of the transfer function is reduced in the frequency domain:
\begin{equation*}
	\min_{G_r(s)} \sum_{k=1}^{N} \left\| G(j\omega_k) - G_r(j\omega_k) \right\|^2
\end{equation*}
In the fitting process, frequency-domain system identification techniques can be employed, such as least-squares method and vector fitting. 
In \cite{8556026}, the authors use a Loewner-based realization of a transfer function:
\begin{equation*}
	H(s) = W (M - sL)^{-1} F
\end{equation*}
where $L$ and $M$ are the Loewner matrix and the shifted Loewner matrix, and $F$ and $W$ are the data matrices, constructed from the frequency response data $s_k, H(s_k)$, partitioned into two sets. The model is later reduced using the singular value decomposition (SVD) of the Loewner matrix:
\begin{equation*}
	L = \Phi \Sigma \Psi^H
\end{equation*}
and truncating directions corresponding to low singular values. The method is applied for the MOR of a large-scale power system and system identification of the black-box power system model.

\subsection{Right-tangential interpolation}
Right-tangential interpolation is employed in \cite{11254892} to develop a reduced-order wind farm model:
\begin{equation*}
	G(\sigma_i)l_i=G_r(\sigma_i)l_i=M_i
\end{equation*}
where $\sigma_i$ are the interpolation frequencies, $l_i$ is the right tangential direction vector, and $M_i$ is the corresponding tangential moment. The tangential moments are calculated directly from the input-output system data organized in the Hankel matrices:
\begin{equation*}
	H_{\ell}(X)=
	\begin{bmatrix}
		x_{0} & x_{1} & \cdots & x_{T-\ell} \\
		x_{1} & x_{2} & \cdots & x_{T-\ell+1} \\
		\vdots & \vdots & \ddots & \vdots \\
		x_{\ell} & x_{\ell+1} & \cdots & x_{T}
	\end{bmatrix}
\end{equation*}
where $X$ is the input or output system data, $l$ is called the Hankel depth, and $T$ is the number of time stamps. This data-driven method, theoretically, is closely linked to the Krylov subspace method.

Although not dealing with converter-dominated power systems, a noteworthy methodology is presented in \cite{RAFIQ2022107908} combining intrusive nonlinear moment-matching and non-intrusive DMD. First, by using signal generator and solving the underlying nonlinear Sylvester equation, the number of states is reduced. Subsequently, the computation cost of the obtained model is reduced by applying DMD through Koopman-based representations.

\section{ML-based dynamic representation}\label{sec.MLDR}

ML techniques can be regarded as MOR methods when they construct a low-dimensional latent representation that captures the essential dynamics of a high-dimensional system. Such approaches can be applied to white-box models, where the governing equations are available, as well as to black-box systems, where only input--output measurements are accessible.

In contrast, if an ML model is trained solely to reproduce the input--output mapping without identifying an underlying low-dimensional dynamical representation, it is more appropriately classified as a surrogate model or dynamic equivalent rather than a reduced-order model in the classical sense. Nevertheless, the distinction between MOR and surrogate modeling is not always clear-cut, and many recent approaches combine elements of both paradigms.

\subsection{ML-based MOR}

Deep learning techniques are leveraged in \cite{Hamid2023Deep}. More precisely, the combination of an auto-encoder and long short-term memory (LSTM) neural network (NN) are used to construct and simulate a reduced-order model. The methodology is applied to the IEEE 118-bus and the European HV transmission network with 1354 buses and shows superior behavior to linear projection-based techniques, such as the POD method. However, the test system does not include PEDs.

Neural ordinary differential equations (Neural ODEs) are employed in \cite{10990040} to reduce a renewable-rich power system, including grid-following and grid-forming converters. To successfully integrate the developed reduced-order model into time-domain simulations with the rest of the power system, an implicit deep equilibrium layer ensures the power system's steady-state equilibrium condition. 
Although the methodology presented in \cite{10298789} does not consider inverter-based generation, it introduces the concept of physics-aware neural ODEs for constructing dynamic equivalents of external power systems. The physical behavior of the unreduced internal system is incorporated into the training process by minimizing both the external-system and internal-system prediction errors, thereby improving the physical consistency of the learned model.

The approach was subsequently extended in \cite{10559616} to account for the limited observability of the internal system through the introduction of a driving-port neural dynamic equivalent. In this more realistic setting, the external model is trained using only the voltages and currents measured at the boundary buses. To compensate for the reduced amount of available information, a recurrent neural network (RNN) is employed to exploit the temporal correlations contained in historical boundary-bus measurements.

A two-machine classical aggregation model of a wind farm, also considering LVRT capabilities, is assembled in \cite{10539343}, enabling reliable transient simulations. The aggregation error dynamics are evaluated using the physics-informed partial differential equations functional identification of nonlinear dynamics (PDE-FIND) algorithm to identify analytical correction equations. Compared to other NN approaches, this algorithm requires less data and possesses physical interpretability.

\subsection{ML-based dynamic equivalents}

A reduced-order model of the external system is developed in \cite{sym17071023} using Deep Neural Networks (DNNs) and boundary measurements between the external and internal systems. An RNN is used for the black-box modeling of the microgrid in \cite{9770523}, inspired by its equivalence to the DAE power system model. Furthermore, layers of LSTMs are also applied in the NN architecture to adapt the microgrid black-box model to a wide operating range and various fault scenarios. Aththanayake et al. in \cite{10407684} used an LSTM-based RNN to develop a dynamic equivalent model of an external system based on the tie-line active and reactive boundary measurements. ML surrogates of wind power and PV power plant, and grid-forming BESS are developed in \cite{Yang2025}, consisting of an LSTM, a cross-layer to capture nonlinear interactions, and the inverter dynamic model to preserve physical constraints and limits. Authors in \cite{10032066} used LSTMs for the error compensation between the real data and the aggregated cyber-physical model of DFIG-based wind farm. 

Li et al. \cite{10633851} employed RNNs with Gate Recurrent Units for black-box modeling of microgrids with inverter-interfaced DERs and a superconducting magnetic energy storage. In addition to using the steady-state samples, the authors showed that data from transient and fault scenarios significantly improves modeling accuracy. Nonlinear autoregressive model with exogenous inputs RNNs is employed in \cite{11168940} to model active and reactive powers of a PV system with voltage support functionality, using historical measurements of powers, irradiance, temperature, and a PCC voltage.

Machine learning algorithms are also employed in combination with other methods. For example, feedforward NNs are employed in \cite{11197966} to compensate for residual errors between the original model and its reduced-order approximation.

\section{Discussion and future perspectives}

The literature overview shows a high variety of different methodologies. Depending on the desired properties of a reduced-order model, the advantages of different techniques are exploited. There is no single method that would fit all the objectives.

Modal reduction’s key advantage is the physical interpretability of the reduced-order model, essential in many power system studies, such as small-signal analysis and transient stability. However, it lacks strong guarantees for stability and error bounds. In contrast, balanced reduction provides theoretical error limits and preserves stability, making it suitable for controller design due to its input-output focus. Yet, it becomes computationally inefficient for large power system models because it requires solving Lyapunov equations. Krylov subspace methods offer scalability for large models, but their error bounds and stability guarantees are harder to evaluate. Data-driven methods, including ML algorithms, accommodate highly nonlinear, complex systems and remain applicable even when the full system model is unavailable; however, their performance depends heavily on the selection and quality of training data.

The state-of-the-art research shows a general maturity of MOR methodologies for conventional $dq$-based power system models. Many proposed solutions successfully address the first enumerated problem (P1) in Section \ref{sec2.2}, related to the multi-timescale nature of PED-dominated power systems. However, there is only a limited contribution to the second problem (P2) of potential converter-stability issues that might occur at supersynchronous frequencies. To handle this, the focus shifts from the MOR methodologies to the fidelity of full-order models. The converter and line models, employed in most of the studies, have limited validity in the frequency domain.

To capture the PWM or unbalanced dynamics, frameworks like dynamic phasors or the harmonic state-space are indispensable for two-level VSC converter models. These modeling approaches are also necessary for capturing internal MMC multi-frequency dynamics \cite{9201165,9541033}. Similarly, for line modeling, the pi model or even simpler RL model are usually utilized. To account for electrical line resonances occurring at higher frequencies, distributed and frequency-dependent line models need to be employed \cite{9007760}. Implementing these modeling methods, necessary for the analysis of harmonic interactions and multi-frequency dynamics inherent to converter-dominated power systems, leads to a significantly increased size of full-order models. For each retained harmonic component, a separate set of variables is required.

In summary, the application of frequency-lifted approaches in power systems with a high share of converter-interfaced resources potentially leads to immense system models. As power systems are being operated ever closer to the stability margin, fast and reliable security assessments are required. This opens new research avenues for developing efficient MOR methodologies for PED-dominated power systems, with the objective of preserving harmonic interactions and multi-frequency dynamics. 

\section*{CRediT authorship contribution statement}
\textbf{G.G.} Conceptualization, Methodology, Investigation, Visualization, Writing - Original Draft, \textbf{M.F.} Investigation, Literature Survey, \textbf{J.-P.M.} Conceptualization, Methodology, Investigation, Writing - Review \& Editing

\section*{Declaration of competing interest}
The authors declare that they have no known competing financial interests or personal relationships that could have appeared to influence the work reported in this paper.



\bibliographystyle{elsarticle-num-names} 
\bibliography{references}






\end{document}